\documentclass[12pt]{iopart}
\usepackage{iopams}

\usepackage{graphicx}

\def\ket#1{|#1\rangle }
\def\bra#1{\langle#1 | }

\def\punkt{\;\; .}
\def\komma{\;\; ,}
\def\expect#1{\langle#1 \rangle}

\def\w{\omega}
\def\H{{\cal H}}
\def\e{\epsilon}

\def\Tr#1{\textrm{Tr}\left[#1\right]}
\def\non{\nonumber\\ }

\begin{document}

\title[Non-Equilibrium  Green's Functions for Quantum Impurity
Models]{A  Numerical Renormalization Group approach to Non-Equilibrium
    Green's Functions for Quantum Impurity Models}

\author{Frithjof B. Anders}
\address{Institut f\"ur Theoretische Physik, Universit\"at Bremen,
  P.O. Box 330 440, D-28334 Bremen, Germany}

\date{\today}

\ead{anders@itp.uni-bremen.de}

\begin{abstract}
We present a  method for the calculation of dynamical
correlation functions of quantum impurity systems out of equilibrium
using Wilson's numerical renormalization group. Our formulation is based on
a complete basis set of the Wilson chain and embeds the recently
derived algorithm for equilibrium spectral functions. Our method
fulfills the spectral weight conserving sum-rule exactly by
construction. 
A local Coulomb repulsion $U>0$ is switched on at $t=0$, and the
asymptotic steady-state spectral functions are obtained for various
values of $U$ as well as magnetic field strength $H$ and temperature $T$. 
These benchmark tests show excellent agreement between the
time-evolved and the directly calculated equilibrium NRG spectra for
finite $U$. This method could be used for calculating steady-state
non-equilibrium spectral functions at finite bias through 
interacting nano-devices.
\end{abstract}

\maketitle

\section{Introduction}

Understanding the influence of the environment onto the
non-equilibrium dynamics of quantum  systems remains
one of the challenging questions of theoretical physics. 
A finite number of quantum mechanical degrees of freedom -- an orbit,
a spin or a qubit -- interacting with a infinitely large bath of
non-interacting bosons or fermions with a continuous energy spectrum,
represents
a typical class of model examples for  such systems. 

These quantum impurity models appear to be at heart of  a variety of
different physical problems. Traditionally,  they were used to
describe the interaction of magnetic impurities within a metallic
host\cite{Hewson93}   or to investigate the dissipation in quantum
mechanics\cite{Leggett1987}. These models have contributed immensely
to our understanding of the low temperature properties
of single-electron transistors\cite{KastnerSET1992,NatureGoldhaberGordon1998}
and the tunneling spectroscopy of adatoms on
metal surfaces.\cite{Manoharan2000,AgamSchiller2000}  In addition,
within the dynamical mean-field theory\cite{Pruschke95,Georges96} or
its cluster extensions\cite{MaierJarrellPruschkeHettler2005} lattice
models for strongly correlated fermions have been mapped onto quantum
impurity problems embedded in a fictitious, self-consistent
bath.

Many approaches to non-equilibrium are based on the
Kadanoff-Baym~\cite{KadanoffBaym62} and Keldysh~\cite{Keldysh65}
techniques. At some time $t_0=0$ a closed system characterized by a density
operator $\hat \rho_0$ evolves according to
the  Hamiltonian $\H(t)$. The immense difficulty of treating the
real-time dynamics of quantum impurity systems stems from the need to
track the full time evolution of the density operator of the entire
system --- environment plus impurity. The Kadanoff-Baym
and Keldysh techniques~\cite{Keldysh65,KadanoffBaym62}  
provide an elegant platform for perturbative expansions of the density
operator. One of the building blocks of such perturbative expansions
are non-equilibrium Green functions. These non-equilibrium
Green functions also contain information on the transients as well as the
steady-state  which might be reached in the long time limit for a
time-independent Hamiltonian. In general, however, perturbative
approaches are plagued by the infra-red divergences caused by
degeneracies on the impurity, making them inadequate for tackling
the change of ground states of quantum impurity models\cite{Wilson75}.

In this paper, we present a different approach for the calculation of
non-equilibrium Green functions of  quantum impurity problems.
We make use of Wilson's numerical renormalization-group (NRG)
method\cite{Wilson75,BullaCostiPruschke2007} and its recent
extension to non-equilibrium
dynamics\cite{AndersSchiller2005,AndersSchiller2006}. 
Spin-spin  non-equilibrium spectral functions obtained by a NRG
calculations were investigated first by 
Costi about ten years ago in the context of the spin-boson
model\cite{Costi97}. Here, we are interested in the evolution of
fermionic spectral functions. We address this problem with a different
approach using the complete  basis set  of 
the Wilson chain \cite{Wilson75,BullaCostiPruschke2007} derived in the
context of the time-dependent numerical renormalization group
\cite{AndersSchiller2005,AndersSchiller2006} (TD-NRG). It  has
already been successfully applied to derive sum-rule
conserving equilibrium Green
functions\cite{WeichselbaumDelft2007,PetersPruschkeAnders2006}.

We focus on a quantum impurity system characterized by the
thermodynamic density operator $\hat \rho_0 \propto \exp(-\beta\H^i)$ for
times $t'<0$. It evolves with respect to the Hamiltonian $\H^f$ for
times $t'\ge 0$. We will derive a closed analytical formula for any
non-equilibrium Green function  $G(t,t')$ for times $t,t'>0$ given  a
time-independent $\H^f$. In contrary to the equilibrium Green
functions,\cite{PetersPruschkeAnders2006,WeichselbaumDelft2007}
transitions between different energy shells require a
double summation over pairs of Wilson shells $(m,m')$. In
Sec.~\ref{sec:II.B}, we prove that this summation can be casted into a
recursion relation involving two different reduced-density matrices
instead of the single one used in the algorithm for equilibrium Green
functions\cite{PetersPruschkeAnders2006,WeichselbaumDelft2007}. It can
be seen analytically that only one of these two
reduced-density matrices contributes  if $\H^i=\H^f$: the equilibrium
algorithm\cite{PetersPruschkeAnders2006} is recovered. Therefore, the 
presented approach to  non-equilibrium  spectral functions embeds the
equilibrium case\cite{PetersPruschkeAnders2006,WeichselbaumDelft2007}
as well.

We will heavily make use of this algorithm in another publication
\cite{AndersSSnrg2008} on the current-voltage characteristics of
interacting nano-devices. In that paper, we will derive a numerical
renormalization group approach based on scattering
states to describe  current-carrying open quantum systems.
In this formulation, the current at finite bias is determined by
the steady-state non-equilibrium (NEQ) spectral
function\cite{MeirWingreen1992,Hershfield1993,Oguri2007,Doyon2007}   which
depends on the density operator of the full system.
At  finite bias, however, the NEQ density operator is only known analytically
for Hamiltonians which commute with the number operator of left
and right-moving electrons,\cite{Hershfield1993,DoyonAndrei2005}
i.e.~for non-interacting quantum impurities.
This analytically known operator $\hat \rho_0$ must be evolved into
the unknown NEQ density operator $\hat \rho$ after switching on a
finite Coulomb repulsion $U$.

We have used the single impurity Anderson model
(SIAM)\cite{KrishWilWilson80a,KrishWilWilson80b}  for benchmarking 
our algorithm.  We have restricted ourselves to changes of local
parameters of the quantum impurity at $t_0=0$. Consequently, the
system has evolved with respect to the full Hamiltonian $\H^f$. For an
infinitely large bath, it is
expected\cite{KadanoffBaym62,Keldysh65,DoyonAndrei2005} that the initial
$\hat\rho_0\propto \exp(-\beta \H^i)$ evolved into the new
thermodynamic density operator  of the fully interacting problem
described by $\H^f$ for times $t\to\infty$, unless it is prohibited by
some conservation law\cite{DoyonAndrei2005}. This is the basic underlying assumption of the
perturbation theory in the Coulomb interaction
$U$\cite{Yamada1974,Yamada1975,YamadaYoshida1978}.
Therefore, the steady-state spectral function obtained from a
time-evolved density operator should be equivalent to the spectra
obtained directly by an equilibrium NRG
calculation\cite{KrishWilWilson80a,KrishWilWilson80b,PetersPruschkeAnders2006,WeichselbaumDelft2007}.

We will use this comparison between both spectra as benchmark for our
algorithm in Sec.~\ref{sec:results}. We will demonstrate  excellent
agreement between these differently calculated spectral functions for
switching on the local Coulomb repulsion $U$ from $U=0$ to a finite
value at various temperatures and local magnetic fields.

\section{Theory}
\label{sec:theory}

Interacting quantum dots, molecular junctions or other nano-devices
are modelled by the interacting region $\H_{imp}$, a set of
non-interacting reservoirs $\H_{bath}$ and a coupling between both
sub-systems $\H_I$
\begin{equation}
  \H = \H_{imp} + \H_{bath} + \H_{I}
\punkt
\end{equation}
We assume that the system is in equilibrium at times $t<0$, and its
properties are determined by the density operator $\hat \rho_0$. One possible
choice  would be $\H_I=0$, which is usually the starting point of
perturbative approaches based on the Keldysh
formalism\cite{Keldysh65}. However, this is not required by our
method. We only demand that the initial density operator can be 
cast in the form
$\hat \rho_0 = \exp(-\beta \H^i)/Z$, where $\H^i$ can be the initial
Hamiltonian of the system in thermodynamic equilibrium for times $t<0$.

At $t_0=0$, we
suddenly switch from the Hamiltonian $\H =\H^i$ to $\H=\H^f$. The retarded
two-time Green function,
\begin{eqnarray}
  G^r_{A,B}(t,t') &= & - i \Tr{\hat \rho_0 [ \hat A(t+t'), \hat B(t') ]_s } \Theta(t)
\nonumber
  \\
 &= & - i \Tr{ \hat \rho_0(t') [ \hat A(t), \hat B ]_s } \Theta(t),
\label{eqn:3}  
\end{eqnarray}
contains information on the correlated dynamics of two operators $\hat
A$ and $\hat B$, where
\begin{eqnarray}
  \hat \rho(t) &=&  e^{-i\H^f t}\hat \rho_0 e^{i\H^f t} \\
  \hat O(t) &=& e^{i\H^f t} O e^{-i\H^f t} 
\punkt
\end{eqnarray}

For fermionic operators the anti-commutator is used
for $[ \hat A(t), \hat B ]_s$ while for Bosonic operators $[ \hat A(t), \hat B ]_s$
represents a  commutator. Eq.~(\ref{eqn:3}) indicates that we can
interpret such a two-time Green function as evolving the density
operator of the system from $\tau=0$ to  the time $\tau=t'$, and
calculating the correlation function of $\hat B$ and $\hat A$ with
respect to the relative time $t>0$. We expect that when changes are
restricted to the local part of the Hamiltonian, i.~e.~$\H_{imp} + \H_I$,
a steady-state or even a new thermodynamic equilibrium
\cite{KadanoffBaym62,Keldysh65,Hershfield1993,DoyonAndrei2005} is
reached for times larger than the largest characteristic time-scale of
the system. In these cases, the limit
\begin{eqnarray}
  \hat \rho_{\infty} = \lim_{t'\to\infty} \hat \rho(t')
\label{eqn:rho-st}
\end{eqnarray}
exists. Eq (\ref{eqn:3}) becomes independent of $t'$, and  $G(t,t')$ only
depends of the relative time $t$ in the steady-state limit.

\subsection{Complete Basis Set }

Wilson's  numerical renormalization group (NRG) method is 
a very powerful tool for accurately calculating  equilibrium properties of 
quantum impurity models. Originally developed for treating the
single-channel, single-impurity Kondo 
Hamiltonian\cite{Kondo62,Wilson75}, this non-perturbative approach
was successfully extended to the Anderson impurity
model\cite{KrishWilWilson80a,KrishWilWilson80b}, and to the two-channel
Anderson\cite{AndersTCSIAM2005} and Kondo
Hamiltonians\cite{Cragg_et_al,PangCox91}.  Recently, it was extended to
equilibrium properties of  impurity models with a {\em bosonic} bath
\cite{BullaBoson2003,BullaVoita2005}, non-equilibrium dynamics of the
spin-boson model \cite{AndersSchiller2006,AndersBullaVojta2007}
or even combinations of both fermionic and Bosonic
baths\cite{GlossopIngersent2005}.

At the heart of this approach is a logarithmic discretization of the
continuous bath, controlled by the discretization parameter $\Lambda >
1$; the continuum limit is recovered for $\Lambda \to 1$. Using an
appropriate unitary transformation,\cite{Wilson75} the Hamiltonian is
mapped onto a semi-infinite chain, defined by a sequence of
finite-size Hamiltonians $\H_m$  with the impurity coupled to the open
end. The iterations are terminated at a finite value of $m=N$ which
defines the Wilson chain of finite length $N$. The finite-size
Hamiltonian $\H_m$ act only on the first $m$ chain links of the Wilson
chain. The length $N$ also determines the temperature $T_N\propto
\Lambda^{-N/2}$ for which the spectral functions are calculated. For a
detailed review on this method see Ref.~\cite{BullaCostiPruschke2007}.

Recently, a {\em complete basis set } for such  a Wilson chain of
length $N$ has been identified\cite{AndersSchiller2005,AndersSchiller2006}.
The set of  eigenstates of  ${\cal H}_m$ can be formally constructed from the
complete basis set $\{ \ket{\alpha_{imp},\alpha_0,\cdots, \alpha_{N} }
\}$ of the NRG chain of length $N$ where the $\alpha_i$
label the configurations on each chain link $i$. Since ${\cal H}_m$
does not act on the chain links $m+1,\cdots,N$, an eigenstate
$\ket{r}$ is written as $\ket{r,e;m}$  where the ``environment''
variable $e = \{\alpha_{m+1}, \cdots,\alpha_N\}$ encodes the
$N - m$ site labels $\alpha_{m+1}, \cdots, \alpha_N$.
The index $m$ is used in this notation to record where the chain
is partitioned into a ``subsystem'' and an ``environment''. After each
iteration the eigenstates of $\H_m$ states are divided in ``discarded'' and
$N_s$ ``kept'' states. The standard NRG proceeds to next iteration $m+1$ using
only the kept states. It
was proven \cite{AndersSchiller2005,AndersSchiller2006} that 
the discarded states from all NRG iterations, i.e $\{ \ket{l,e;m}_{dis} \}$
also form a complete basis set. Regarding all eigenstates of the final
NRG iteration as discarded, one can formally write the
 Fock space of the $N$-site chain in the form
${\cal F}_{N} = {\rm span} \{|l,e;m \rangle_{dis}\}$, and
the following completeness relation holds:
\begin{eqnarray}
\sum_{m = m_{\rm min}}^N \sum_{l,e}
          \ket{l,e;m}_{dis}\ _{dis}\bra{l,e;m} &=& 1 \; .
\label{equ:complete-basis}
\end{eqnarray}
Here the summation over $m$ starts from the first iteration
$m_{\rm min}$ at which a basis-set reduction is imposed.
All traces  below will be carried out with respect to
this basis set. Hence, the evaluation of the spectral functions
will not involve any truncation error. Note also that we
made no reference to a particular Hamiltonian ${\cal H}$ 
in constructing the basis set $\{ |l,e;m \rangle_{dis} \}$.

At each iteration $m$, the Fock space ${\cal F}_{N}$ of a Wilson chain
with fixed length $N$ is partitioned by
all previously discarded states 
\begin{eqnarray}
 1_m^- &=& \sum_{m'=m_{min}}^{m-1} \sum_{l',e'}
               \ket{l',e';m'}_{dis}\ _{dis}\bra{l',e';m'} \; ,
\label{eqn:partition-fock-space-i}
\end{eqnarray}
and all states present $r$ at iteration $m$
\begin{eqnarray}
1_m^+ &=& \sum_{m'= m}^N \sum_{l',e'}
               \ket{l',e';m'}_{dis}\ _{dis}\bra{l',e';m'} \; .
\nonumber \\
& =& \sum_{r,e} \ket{r,e;m} \bra{r,e;m} \; .
\label{eqn:8}
 \end{eqnarray}
We will make extensive use of the completeness relation 
\begin{eqnarray}
  1 &=& 1_m^- + 1_m^+ 
\label{eqn:completness}
\end{eqnarray}
in the following section.

\subsection{Derivation of the NRG non-equilibrium Green function}
\label{sec:II.B}

For the moment, we will consider only the first term of the commutator
of the retarded Green function $I(t',t) = \Tr{\hat \rho(t') \hat A(t)
  \hat B}$.  If the operator $\hat O_t = \hat A(t)  \hat B$ were a ``local''
operator, i.e.~an operator which only acts on impurity  degrees of
freedom or a Wilson chain of length $m_{min}$ up to which all
states are still maintained, we could use the
TD-NRG\cite{AndersSchiller2005,AndersSchiller2006} to 
calculate the time evolution of $O_t(t') = \Tr{\hat \rho(t') \hat O_t}$.

In general, the time evolution of a local operator $\hat O$ 
leads to an operator $\hat O_t$ which acts on all chain degrees
of freedom. Each operator  $\hat O_t$ can always be expanded in
outer products 
of all  many-body states spanning the Fock-space. Here, we will
restrict  ourselves always to a many-body Fock-space basis which is an
approximate eigenbasis of the Wilson chain Hamiltonian. For the
application of the TD-NRG, we require that  the matrix elements of $\hat O_t$ 
remain diagonal in and independent of  the environment degrees of freedom $e,e'$
\begin{eqnarray}
   \langle r,e;m|\hat O_t | s,e';m \rangle &=&   \delta_{e,e'}
   O^m_{rs}(t)
\ .
\label{eqn-10}
\end{eqnarray}
Then  the operator qualifies  as local operator as defined in
Eqn.~(21) of Ref.~\cite{AndersSchiller2006}.
We insert the
completeness relation 
Eq.~(\ref{eqn:completness}) between $\hat A(t)$ and $\hat B$ and
obtain the two
contributions 
\begin{eqnarray}
\label{eqn:local-operator-partitioning}
&&  \langle r,e;m|\hat A(t)  \hat B | s,e';m \rangle =  \langle r,e;m|\hat A(t)(1_m^+ + 1_m^-) \hat B | s,e';m
  \rangle
\nonumber \\
&=&
\sum_{k,e''} \langle r,e;m|\hat A(t)| k,e'';m
  \rangle\langle k,e'';m|  \hat B | s,e';m \rangle
 \\
& & +\sum_{m''=m_{min}}^{m-1} \sum_{l'',e''}
\langle r,e;m|\hat A(t) \ket{l'',e'';m''}_{dis}\ _{dis}\bra{l'',e'';m''} 
\hat B | s,e';m \rangle
\nonumber
\punkt
\end{eqnarray}
Restricting the operators $\hat A$ and $\hat B$ to local operators,
the first term remains diagonal in $e,e'$\cite{AndersSchiller2006}. In the
second term, we again make use  of Eq.~(\ref{eqn:completness}), but
partitioning the Fock-space of the Wilson chain with respect to
iteration $m''$:
\begin{eqnarray}
&&  \langle r,e;m|(1_{m''}^+ + 1_{m''}^-)\hat A(t) 
\ket{l'',e'';m''}_{dis}\ _{dis}\bra{l'',e'';m''} 
\hat B (1_{m''}^+ + 1_{m''}^-)| s,e';m \rangle 
\nonumber \\
&&=  \langle r,e;m|1_{m''}^+\hat A(t) 
\ket{l'',e'';m''}_{dis}\ _{dis}\bra{l'',e'';m''} 
\hat B \;1_{m''}^+ | s,e';m \rangle 
\nonumber \\
&&= 
\sum_{k_1,e_1} \sum_{k_2,e_2}
\langle r,e;m|   k_1,e_1;m'' \rangle 
\langle  k_1,e_1;m''|\hat A(t) 
\ket{l'',e'';m''}_{dis}\
\nonumber \\
&&
\phantom{\sum_{k_1,e_1} \sum_{k_2,e_2}}
\times
 _{dis}\bra{l'',e'';m''} \hat B |   k_2,e_2;m''
\rangle 
\langle  k_2,e_2;m''|  s,e';m \rangle 
\nonumber \\
&&
= \sum_{k_1,e_1} \sum_{k_2,e_2}
\langle r,e;m|   k_1,e_1;m'' \rangle 
A^{m''}_{k_1,l''}e^{i(E^{m''}_{k_1}-E^{m''}_{l''})t} \delta_{e_1,e''}
\nonumber \\
&&
\phantom{\sum_{k_1,e_1} \sum_{k_2,e_2}}
\times
B^{m''}_{l'',k_2}\delta_{e_2,e''}
\langle  k_2,e_2;m''|  s,e';m \rangle 
\,\; .
\label{eqn:local-operator}
\end{eqnarray}
Note that $1^-_{m''}| s,e';m \rangle=0$ holds for $m''<m$, and the indices $k_1$ and $k_2$ include all states present
at iteration $m''$ as seen from the definition of $1^+_{m''}$ in
Eq.~(\ref{eqn:8}). The locality of the operators  $\hat A$ and
$\hat B$ has been used and leads to the condition $e_1=e_2$. Since
$m''<m$, we can partition the environment 
degrees of freedom $e_1$ into $e_1=(\tilde e_1, e'_1)$ where
$e'_1$ labels the Wilson chain degree of freedom starting from chain
link $m+1$. We obtain only non-zero matrix elements $\langle r,e;m|
k_1,e_1;m'' \rangle \langle  k_2,e_1;m''|  s,e';m \rangle $, if
$e=e'_1=e'$. Therefore, Eq.~(\ref{eqn-10}) holds, 
and the matrix elements in Eq.(\ref{eqn:local-operator}) are
independent of $e$.

Consequently, the operator $\hat O_t = \hat
A(t)  \hat B$ qualifies as a local operator in the sense of the
TD-NRG\cite{AndersSchiller2005,AndersSchiller2006} for each time $t$,
and $I(t',t)$ is given by the fundamental equation of the TD-NRG,
Eq.~(3) in Ref.~\cite{AndersSchiller2005}, 
\begin{eqnarray}
I(t',t) &=&  \sum_{m = m_{min}}^{N}\sum_{r,s}^{trun} \;
        e^{i(E_{r}^m -E_{s}^m)t'}
        O_{r,s}^m(t) \rho^{red}_{s,r}(m) \; .
\nonumber \\
\label{eqn:time-evolution}
\end{eqnarray}
Here $O_{r,s}^m(t) = \langle r,e;m|\hat A(t)  \hat B | s,e;m \rangle$
is independent of $e$, and 
reduced density matrix  $\rho^{red}_{s,r}(m)$ 
\begin{equation}
\rho^{red}_{s,r}(m) = \sum_{e}
          \langle s,e;m|\hat{\rho}_{0} |r,e;m \rangle 
\label{eqn:reduced-dm-def}
\end{equation}
is given in the NRG basis of $\H^f$. At each time $t'$, the spectral
information is encoded in the time evolution of $\hat O(t)$.

Inserting Eq.~(\ref{eqn:local-operator-partitioning}) into
Eq.~(\ref{eqn:time-evolution}) yields two terms. The first
contribution to $I(t,t_2)$ remains diagonal in the 
iteration index $m$ and is given by the following expression
\begin{eqnarray}
  I_1(t',t) &=& \sum_{m = m_{min}}^{N}\sum_{r,s}^{trun} \sum_{k} \;
  e^{i(E_{r}^m -E_{s}^m)t'} A^m_{r,k}e^{i(E_{r}^m -E_{k}^m)t}
  \nonumber \\
  && \times     B^m_{k,s}\rho^{red}_{s,r}(m)  
\punkt
\end{eqnarray}
The restricted sum $\sum_{r,s}^{trun}$ requires that at
least one of those indices $r,s$  labels a discarded state at
iteration $m$.
The second contribution to $I(t',t)= I_1(t',t)+
I_2(t',t)$, $I_2(t',t)$, contains a double summation  over the
iteration indices $m$ and $m''$ 
\begin{eqnarray}
   I_2(t',t) &=&  \sum_{m = m_{min}}^{N}\sum_{r,s}^{trun} 
\sum_{m''=m_{min}}^{m-1} \sum_{e}   e^{i(E_{r}^m -E_{s}^m)t'}
\nonumber \\
&& \times 
\sum_{l'',e''}
\langle r,e;m|\hat A(t) \ket{l'',e'';m''}_{dis} 
\nonumber \\
&& \times 
\ _{dis}\bra{l'',e'';m''} 
\hat B | s,e;m \rangle
\nonumber \\
&& \times 
\bra{s,e;m}\hat\rho_0\ket{r,e;m}
\label{eqn:secon-part-I2}
\end{eqnarray}
which prevents a simple evaluation of the matrix elements of $\hat A$ and
$\hat B$. 
Now, we insert Eq.~(\ref{eqn:local-operator}) into
Eq.~(\ref{eqn:secon-part-I2}) 
and arrive at
\begin{eqnarray}
   I_2(t',t) &=&  \sum_{m = m_{min}}^{N}\sum_{r,s}^{trun} 
\sum_{m''=m_{min}}^{m-1} 
 \sum_{k_1,k_2}
  e^{i(E_{r}^m -E_{s}^m)t'}
\sum_{l'',e''}
A^{m''}_{k_1,l''}e^{i(E^{m''}_{k_1}-E^{m''}_{l''})t}
B^{m''}_{l'',k_2}
\nonumber \\
&& \times 
\sum_{e,e_2} 
\langle r,e;m|   k_1,e_2;m'' \rangle 
\bra{s,e;m}\hat\rho_0\ket{r,e;m}
\langle  k_2,e_2;m''|  s,e;m \rangle 
\, .
\nonumber \\
\end{eqnarray}
The summation $\sum_{m = m_{min}}^{N}$ and 
$\sum_{m''=m_{min}}^{m-1}$ implies that $m''<m$. Therefore, the
summation can be arranged to 
\begin{eqnarray}
   I_2(t',t) &=&  
\sum_{m'' = m_{min}}^{N-1}
\sum_{l''}^{trun}  \sum_{k_1}
\sum_{k_2} A^{m''}_{k_1,l''}(t)
B^{m''}_{l'',k_2}
\nonumber \\
&& \times 
\tilde \rho^{red}_{k_2,k_1}(m'',t') \; ,
\end{eqnarray}
where the indices $k_1,k_2$ run over all eigenstates of $\H_{m''}$
present at iteration $m''$, but the index $l''$
remains restricted to the discarded states. 
%
%
In the last step, we have defined a second
reduced density matrix $\tilde \rho_{k_1,k_2}(m'',t')$ as
\begin{eqnarray}
  \label{eq:reduced-rho-a}
\tilde \rho_{k_2,k_1}(m'',t') &=& \sum_{m=m''+1}^{N} 
\sum_{r,s}^{trun} 
\sum_{e,e_1}
\langle r,e;m \ket{k_1,e_1; m''}  
\nonumber \\
&& 
\times 
\bra{k_2,e_1;m''} s,e;m \rangle
\bra{s,e;m}\hat\rho_0\ket{r,e;m}
\nonumber \\
&& \times 
e^{i(E_{r}^m -E_{s}^m)t'}
\punkt
\end{eqnarray}
Partitioning the environment variable $e_1$ into
$e_1=(\alpha_{m''+1},\cdots, \alpha_{m},e')$, the relation
\begin{eqnarray}
  \label{eq:reduced-rho-ii}
\tilde \rho^{red}_{k_2,k_1}(m'',t') &=& \sum_{m=m''+1}^{N} 
\sum_{r,s}^{trun} 
\sum_{\{\alpha_i\}}
\rho^{red}_{s,r}(m) e^{i(E_{r}^m -E_{s}^m)t'}
\nonumber \\
&& \times 
\langle r;m \ket{k_1,\{\alpha_i\}; m''}  
\bra{k_2,\{\alpha_i\};m''} s;m \rangle \;\; 
\nonumber \\
&& 
\end{eqnarray}
is obtained. Here, we explicitly made use of the fact that the matrix
elements $\bra{k_2,e_1;m''} s,e;m \rangle$ are diagonal in $e'$ and
$e$ and independent of $e$. The summation over $e$ only enters the
definition of $\rho^{red}_{s,r}(m)$.

%
Eq.~(\ref{eq:reduced-rho-ii}) connects  $\tilde
\rho^{red}_{k_2,k_1}(m,t')$ to all reduced density operators
$\rho^{red}_{s,r}(m')$ from the later iterations $m'>m$. If  $\tilde
\rho^{red}_{k_2,k_1}(m+1,t')$ is given,  $\tilde \rho_{k_2,k_1}(m,t')$ obeys the
following recursion relation
\begin{eqnarray}
  \label{eq:reduced-rho-ii-recurs}
 \tilde \rho^{red}_{k_2,k_1}(m,t') &=& 
\sum_{r,s}^{trun} 
\sum_{\alpha_{m+1}}
\bra{k_2,\alpha_{m+1};m} s;m+1 \rangle
\nonumber \\
&& \times
\left[\rho^{red}_{s,r}(m+1) e^{i(E_{r}^{m+1} -E_{s}^{m+1})t'}\right]
\langle r;m+1 \ket{k_1,\alpha_{m+1}; m}  
\nonumber \\
&& +
\sum_{k',k''}^{trun} 
\sum_{\alpha_{m+1}}
\bra{k_2,\alpha_{m+1};m} k';m+1 \rangle
\nonumber \\
&& \times
 \tilde \rho^{red}_{k',k''}(m+1,t')
\langle k'';m+1 \ket{k_1,\alpha_{m+1}; m}  
\punkt
\end{eqnarray}
which we have obtained  from Eq.~(\ref{eq:reduced-rho-ii}). We
initialize this recursion with  $\tilde \rho^{red}_{k',k''}(N,t')=0$.
Defining the auxiliary matrix
\begin{eqnarray}
 \label{eqn:init-rho-prime}
   \rho'_{r,s}(m+1,t')&=&\rho^{red}_{s,r}(m+1) e^{i(E_{r}^{m+1}
     -E_{s}^{m+1})t'} + \tilde \rho_{k',k''}(m+1,t')
\;\; ,
 \end{eqnarray}
the recursion relation (\ref{eq:reduced-rho-ii-recurs}) has the same
structure as Eq.~(40) of Ref.~\cite{AndersSchiller2006}.

Note that the overlap matrix elements
$\bra{k_2,\alpha_{m+1};m} k';m+1 \rangle$ are identical to the matrix elements
$A_{k',k2}^{\alpha_{m+1}}$ as defined in Eq.~(2) of
Ref.~\cite{WeichselbaumDelft2007}.
Matrix elements of this type $\langle r;m \ket{k_1,\{\alpha_i\};
  m''}$ can be evaluated directly using a product of $m-m''$
such $A$-matrices \cite{WeichselbaumDelft2007}.

At each recursion step  $\rho'_{r,s}(m+1,t')$
involves two terms which 
contribute  matrix elements to  different sectors of $\rho'$. By
construction, $\tilde \rho_{k',k''}(m+1,t)$ has  
only non-zero matrix elements for $k'$ and $k''$ being both 
retained states of the NRG iteration $m+1$. 

The restricted sum over $r$ and $s $
projects out the
other sectors of the  matrix $\rho'_{r,s}(t')=\rho^{red}_{s,r}(m+1,t') + \tilde
\rho_{s,r}(m+1,t')$ for which at least one of the indices $s,r$ labels a
discarded state. Instead of a single reduced density matrix, we 
 need to keep tract of two matrices at each iteration, namely
$\rho^{red}_{r,s}(m)$ and $\tilde \rho^{red}_{r,s}(m,t')$.

Then,  the two contributions to $I(t',t)$ read
\begin{eqnarray}
  I(t',t) &=& \sum_{m = m_{min}}^{N}\sum_{r,s}^{trun} \sum_{k} \;
  e^{i(E_{r}^m -E_{s}^m)t'} A^m_{r,k}e^{i(E_{r}^m -E_{k}^m)t}
  \nonumber \\
  && \times 
    B^m_{k,s}\rho^{red}_{s,r}(m)  
  \nonumber \\
  && +
  \sum_{m = m_{min}}^{N-1}
\sum_{l'}^{trun}  \sum_{k_1}
\sum_{k_2} A^{m}_{k_1,l'}e^{i(E_{k_1}^m -E_{l'}^m)t}
B^{m}_{l',k_2}
\nonumber \\
&& \times 
\tilde \rho^{red}_{k_2,k_1}(m,t')
\punkt
\label{eqn:19}
\end{eqnarray}
This formally requires only a single  summation over $m$: the
second summation over $m'$  has been absorbed into the definition of $\tilde
\rho^{red}_{k_2,k_1}(m,t')$. Note that the index $l'$ labels all
discarded states at iteration $m$. Obviously, the same type of
calculation must also 
be performed for the second term of the commutator in Eq.~(\ref{eqn:3})
in order to obtain all contributions for the Green function. Fourier
transformation of  Eq.~(\ref{eqn:19}) with respect to $t$  yields the spectral information
of interest.

It has to be emphasized that only energetic approximations have been
made. The NRG truncation influences the  partitioning of the
states, but the completeness of the basis is always
guaranteed\cite{AndersSchiller2005,AndersSchiller2006}.  Therefore,
the spectral sum-rule remains fulfilled exactly for each time $t'$ as in
the equilibrium case\cite{PetersPruschkeAnders2006}. It is straight
forward to apply our algorithm also to  the lesser and greater Green
functions $G^<(t,t')$ and $G^>(t,t')$ as discussed in 
Ref.~\cite{WeichselbaumDelft2007}.

\subsection{Steady-state limit}
\label{sec:II.C}

For all systems in which a time-independent steady-state density
operator $\hat \rho_{\infty}$ is reached, Eq.~(\ref{eqn:rho-st})  becomes 
equivalent  to
\begin{eqnarray}
  \hat \rho_{\infty} = \lim_{T\to\infty} \frac{1}{T} \int_0^T d\tau \hat \rho_0(\tau)
\label{eqn:rho-st-ave}
\punkt
\end{eqnarray}
This formulation is particularly useful for a discretized
representation of an infinitely large system since artificial finite
size oscillations are averaged out. The steady-state limit of the 
two-time Green function,
\begin{eqnarray}
G^r_{\infty}(t) &=& \lim_{t'\to\infty}  G^r_{A,B}(t,t') \nonumber
\\
&=& - i \Tr{ \hat \rho_{\infty} [ A(t), B ]_s } \Theta(t),
\label{eqn:gf-st}  
\end{eqnarray}
is obtained using  Eqs.~(\ref{eqn:19}) and
(\ref{eqn:rho-st-ave}) by noting that
\begin{equation}
 \lim_{T\to\infty} \frac{1}{T} \int_0^T d\tau  e^{i(E_{r}^m -E_{s}^m)\tau}
= \delta_{E_r,E_s}
\punkt  
\end{equation}
In the first part of Eq.~(\ref{eqn:19}) as well as in the recursion relation
(\ref{eq:reduced-rho-ii-recurs}), the reduced density matrix
$\rho^{red}_{s,r}(m)$ contributes only energy diagonal
matrix elements. In general, however, the reduced density matrix
$\tilde \rho^{red}_{k,k'}$ will not be diagonal in the NRG eigenbasis.

We introduce the integral   $L_{A,B}(t')$ of the Fourier transformed  Green
function $ G^r_{A,B}(\w,t')$ with respect to $t$ as 
\begin{equation}
  L_{A,B}(t') =  -  \int_{-\infty}^\infty \frac{d\omega}{\pi} \, 
 \Im m  G^r_{A,B}(\w,t') \;\; .
\end{equation}
For operators $\hat A$ and $\hat B$, whose anti-commutator --
commutator for bosonic operators -- remains
constant, $L_{A,B}(t')$ defines a sum-rule independent of $t'$ which
is fulfilled exactly 
by our approach at any time $t'$ due to the usage of a complete basis set.
Therefore,  the averaged sum-rule 
 \begin{equation}
   L_{A,B} =  \lim_{t'\to\infty} L_{A,B}(t')  = -\lim_{T\to\infty} \frac{1}{T} \int_0^T d\tau
   \int_{-\infty}^\infty \frac{d\omega}{\pi} \, 
 \Im m  G^r_{A,B}(\w,\tau)
 \end{equation}
remain exacty fullfilled as well. An example would be the
single-particle spectral function obtained from
Eq.~(\ref{eqn:19}) by setting $A=f_\sigma$ and
$B=f^\dagger_\sigma$. In this case $L_{f_\sigma,f^\dagger_\sigma}(t')
=1$.
In fact, we use this criterion to check
explicitly the sum-rule conservation and found that it remains always within
machine precision with an error of $10^{-15}$ independent of all parameters.

A word is in order about the usage of the term ``steady-state.'' 
We expect that a steady-state is always reached at long times for a
time independent Hamiltonian\cite{DoyonAndrei2005} $\H^f$ in quantum
impurity systems. In a closed but infinite quantum system,
where only $\H_{imp}+\H_{I}$ has been changed, the steady-state
will be identical to the thermodynamic equilibrium described by the density
operator $\hat\rho  = \exp(-\beta \H^f)/Z_f$, in the sense that all
local expectation values calculated with $\rho_\infty$ and  $\hat\rho$
will be the same. It requires that the limit
$\lim_{t'\to\infty}\lim_{V\to\infty}$ is taken such that $t'\ll V$ in
appropriate dimensionless units.

A steady-state rather than a thermodynamic
equilibrium\cite{Hershfield1993,DoyonAndrei2005} 
will be reached for an open quantum system
in the limit $t'\to\infty$ \cite{Hershfield1993,DoyonAndrei2005} at
finite bias.
Again, it requires that the limit
$\lim_{t'\to\infty}\lim_{V\to\infty}$ is taken in the correct order.
However, within a
discretized representation of such a quantum impurity system, we can
never distinguish between the approach to a true thermodynamic
equilibrium and non-equilibrium steady-state for times $t'\to 
\infty$. Therefore, we will always use the term ``steady-state''
throughout the paper even for situations where it can be proven that
the corresponding continuum limit of the model approaches the
thermodynamic limit for infinitely long times\cite{DoyonAndrei2005}.
In fact,  the difference between our steady-state  and
equilibrium spectral function  will serve as a criterion for the
quality of our approach.

\subsection{Recovering the sum-rule conserving equilibrium NRG Green function}
\label{sec:II.D}

Equation  (\ref{eqn:19}) must contain all contributions to the
equilibrium Green function
\cite{WeichselbaumDelft2007,PetersPruschkeAnders2006} as well. In
equilibrium, the initial and final Hamiltonian are identical
$(\H=\H^i=\H^f$), the density operator $\hat \rho_0$ commutes with
$\H$. The overlap matrix $S_{r,s}$ between eigenstates of $\H^i$ and
$\H^f$, $S_{r,s}= \ _{i}\langle s;m|r,m\rangle_{f}$ must be diagonal. Then,
$I_1$ contributes with an energy diagonal $\rho^{red}_{s,r}(N)$ only
on the last Wilson shell and is identical to Eq.~(11) of
Ref.~\cite{PetersPruschkeAnders2006}. For $m <N$,
$\rho^{red}_{s,r}(N)$ has only non-zero matrix elements for $r$ and
$s$ being a kept state, which are explicitly excluded by the
summation restriction. Therefore, $\rho^{red}_{s,r}(m)$  contributes
only once  to the reduced density matrix $\tilde
\rho^{red}_{k_2,k_1}(m,t')$ in the recursion relation
Eq.~(\ref{eq:reduced-rho-ii-recurs}), namely at iteration $N-1$. As a
consequence, the reduced 
density matrix $\tilde \rho^{red}_{k_2,k_1}(m,t')$ becomes time
independent in  equilibrium and equal to the reduced density matrix
$\rho^{red}_{k_2,k_1}(m)$, i.e.~$\tilde \rho^{red}_{k_2,k_1}(m,t') = 
\rho^{red}_{k_2,k_1}(m)$. The Fourier transformation of
$I_2(t'=0,t)$ with respect to $t$ yields  Eq.~(16) of
Ref.~\cite{PetersPruschkeAnders2006}.

\subsection{The non-equilibrium NRG algorithm}

\label{sec:td-nrg-algo}

As in the equilibrium NRG,\cite{Wilson75} each chain length $N$
corresponds to a temperature $T_N \propto \Lambda^{-N/2}$.
For ${\cal H}^i$ and   ${\cal H}^f$, two simultaneous  NRG
runs are performed in order to generate the density operator $\hat\rho_0$
using $\H^i$ and the eigenenergies of $\H^f$ for the time
evolution. At each iteration $m$, we calculated the overlap
matrix $S_{r,r'}(m)$ between all eigenstates $r$ of $H_m^f$ and
all eigenstates $r'$ of   $\H_m^i$\cite{AndersSchiller2006}. This information, as well as the
unitary matrices diagonalizing $\H^i_m$ and $\H^f_m$
are stored. At the end of the NRG runs, the equilibrium density matrix
\cite{Wilson75,BullaCostiPruschke2007,AndersSchiller2005,AndersSchiller2006}
is calculated using the last iteration of
$H_N^i$:
\begin{eqnarray}
  \label{eqn:spectral-rho-0}
  \hat \rho_0 &=& 
  \frac{1}{Z_N} \sum_{l} e^{-\beta_N E^N_l} \ket{l;N}\bra{l;N}
\end{eqnarray}
where $Z_N=  \sum_{l} \exp({-\beta_N E^N_l})$.

We have  implemented the TD-NRG algorithm\cite{AndersSchiller2006}
recursively by going backwards from $m$ to $m-1$. 
For each backward iteration, we perform the following steps:
\begin{enumerate}
\item calculate the reduced density matrix in the basis of $\H_m^i$ using
  using Eq.~(40) in Ref.~\cite{AndersSchiller2006}

\item   calculate  $\rho'_{r,s}(m+1,t')$ according to
  Eq.~(\ref{eqn:init-rho-prime}) 

\item  calculate $\tilde\rho^{red}_{r,s}(m,t')$ using the recursion
  Eq.~(\ref{eq:reduced-rho-ii-recurs})

\item  combine $\tilde\rho^{red}_{r,s}(m,t')$ and  $\rho^{red}_{r,s}(m+1)$
  to a single reduced density matrix

\item  evaluate the contribution of iteration $m$ to the 
  excitation spectrum obtained  by Fourier transform Eq.~(\ref{eqn:19})

\item steps (i)-(v) are repeated until we reach the iteration
  $m_{min}$ at which no state was eliminated.
\end{enumerate}

While the selection of retained states in the NRG run for $\H^i$ is
determined by the density matrix\cite{Wilson75}, the selection of states
of $\H^f$ is guided by the notion of maximizing the overlap with the
eigenstates of $\H^i$. Amongst different truncation 
schemes, which we have implemented, the simplest was the most
effective\cite{AndersSchiller2005,AndersSchiller2006}. In this
truncation scheme, we selected the lowest 
eigenstates of  $\H^f_m$ at the end of each iteration $m$ as well. 

In Ref.~\cite{AndersSSnrg2008} the current through
a nano-device coupled to two leads is investigated as function of the
finite applied bias using the algorithm for NEQ spectral function presented
here. The device is described by  a two-band model. 
Each band representing the bath continuum for either left or
right-moving scattering states will be set to a different
chemical potential $\mu_\alpha$, $\alpha=L,R$. The potential different
$V=\mu_R-\mu_L$ drives a finite current through the nano-device. In
this case, the NRG run for $\H^i$ obtains a faithful many-body
representation of the density operator of the non-interaction problem
($U=0$) 
\begin{eqnarray}
  \rho_0 \propto  e^{-\beta(\H^i -\hat Y_0)}
\end{eqnarray}
where operator\cite{Hershfield1993} 
\begin{eqnarray}
  \hat Y_0 &=& \sum_\alpha \mu_\alpha N_\alpha
\end{eqnarray}
replaces the usual number operator for a grand canonical ensemble in
order to include the different potentials $\mu_\alpha$ of the
scattering states.

After each iteration for $\H_m^f$, one would like to retain the states
with the largest overlap with the eigenstates of $\H^i_m$.  
These eigenstates of $\H^f$ are generally expected to be connected to the
eigenstates of $\H^i$ of the same eigenenergy relative to the ground
state by the Lippmann-Schwinger equation for a model with a 
continuous bath. 
In practice, we select those
eigenstates of $\H^f_m$ which have the lowest diagonal matrix elements
of the operator $\H^f_m-\hat 
Y_0$. Therefore, the eigenenergies $E_s$ of $\H^f_m$  can be divided into two
contributions
\begin{eqnarray}
  E_s &=&\Delta E_s + \sum_\alpha \mu_\alpha n^s_\alpha \; .
\end{eqnarray}
The first term $\Delta E_s$ is of the order $\Lambda^{-m/2}$ due to
the truncation scheme, and the
second term 
is defined by
\begin{eqnarray}
\bra{s}\hat Y_0\ket{s} &=&   \sum_\alpha \mu_\alpha n^s_\alpha = 
\sum_\alpha \mu_\alpha \bra{s}\hat N_\alpha \ket{s} 
\punkt
\end{eqnarray}

The question  of the distribution and magnitude of the 
the excitation energies $\Delta E_{rs}= E^m_r - E^m_s$
entering Eq.~(\ref{eqn:19})  arises in order to understand the
redistribution of spectral weight at finite bias.
$\Delta E_{rs}$ involves eigenenergies of  $\H^f_m$ and
is  given by
\begin{eqnarray}
  \Delta E_{rs} &=& \Delta E_r - \Delta E_s
+ \sum_{\alpha}\mu_\alpha \left(n_\alpha^r-n_\alpha^s\right)
\label{eqn:delta-E-chem}
\end{eqnarray}
The single-particle spectral function is obtained from
Eq.~(\ref{eqn:19}) by setting $A=f_\sigma$ and $B=f^\dagger_\sigma$. Only
those states $r$ and $s$ can  contribute to the  spectral function whose total number of particles differs by exactly
one electron, i.e.~
\begin{eqnarray}
\label{eqn:delta-N}
  \sum_\alpha \left(n_\alpha^r-n_\alpha^s\right) &=&
 \pm 1 \;\; .
\end{eqnarray}
Substituting Eq.~(\ref{eqn:delta-N})
into (\ref{eqn:delta-E-chem}) yields the two equivalent ways of
writing the excitation energies
\begin{eqnarray}
\label{eq:delta-E-mu-L}
   \Delta E_{rs} &=& \Delta E_r - \Delta E_s + (\mu_R-\mu_L) \left(n_R^r-n_R^s\right)
   \pm \mu_L
\\
 &=& \Delta E_r - \Delta E_s + (\mu_L-\mu_R)\left(n_L^r-n_L^s\right)
   \pm \mu_R
\label{eq:delta-E-mu-R}
\punkt
\end{eqnarray}

For models with a channel conservation law,
$|\left(n_\alpha^r-n_\alpha^s\right)|= 0,1$ 
must hold. As a consequence, the  excitation energies  $\Delta E_{rs}$
are centered around the  two chemical potentials $\mu_\alpha$. 
For interacting quantum impurity models which violate channel
conservation\cite{Hershfield1993,AndersSSnrg2008}, the differences  $|\Delta
N_\alpha^{rs}|$ are given arbitrary numbers. By inserting a finite
value of $\left(n_\alpha^r-n_\alpha^s\right)$ 
into Eq.~(\ref{eq:delta-E-mu-L}) or  (\ref{eq:delta-E-mu-R}), it becomes
apparent that the  energy difference $\Delta E_{rs}$ will be shifted
away from either chemical potential by multiples of the chemical
potential differences  $V=\mu_L-\mu_R$\cite{Hershfield1993,AndersSSnrg2008}.


A word is in order concerning the  the frequency resolution. In the
usual equilibrium NRG the lowest resolvable
frequency\cite{BullaCostiPruschke2007} coincides with the temperature
$T_N \propto \Lambda^{-N/2}$ set by the length of the Wilson chain. 
The non-equilibrium Green functions $G(t,t')$  depends on two different times. The
Fourier transformation with respect to relative time $t$  remains meaningful 
even in the  limit $t'\to \infty$, since the steady-state density operator
$\rho_\infty$ exists and is well defined by
Eq.~(\ref{eqn:rho-st-ave}). However, the  smallest excitation energy
resolved might be larger than  $\w_{N} \approx \Lambda^{-N/2}$ due to
the difference between $\rho^{TD-NRG}_\infty$ obtained via
Eq.~(\ref{eqn:rho-st-ave})   and the exact steady-state density
operator for a bath continuum. Depending on the bias $V$ and values of
$U$ the lower boundary for frequency resolution increases to  $\w_{low} \approx \Lambda^{-m/2}$ 
which typically $m=N-1$ to $m=N-3$.  In all cases, we investigated in
Ref.~\cite{AndersSSnrg2008}, the
bias $V$ remains significantly larger that $\w_{low}$.

\section{Results}
\label{sec:results}

\subsection{The single impurity Anderson model}

In order to demonstrate the potential of this approach, we will
present results for the single-particle spectral functions of the
single impurity Anderson model (SIAM) for which the equilibrium
spectral functions are well studied
\cite{CostiHewsonZlatic94,BullaHewsonPruschke98,BullaCostiVollhardt01,BullaCostiPruschke2007,PetersPruschkeAnders2006}
and can serve as benchmarks.

The  Hamiltonian of the
SIAM\cite{Anderson61,KrishWilWilson80a,KrishWilWilson80b} 
 \begin{eqnarray}
   \H &=& \sum_{k\sigma} \e_{k\sigma} c^\dagger_{k\sigma}c_{k\sigma}
+H_{imp}
\non
&& 
 + V\sum_{k\sigma}  \left( c^\dagger_{k\sigma} f_\sigma
 +f^\dagger_{\sigma} c_{k\sigma}\right)
\\
\label{eqn:h-imp}
\label{eqn:25}
\H_{imp} &=& \H_0 + \H_{U} \\
&=&
\sum_{\sigma} \left(\e_f +\frac{U}{2} - \frac{\sigma}{2} H\right)
 f^\dagger_\sigma f_\sigma 
 + \frac{U}{2}\left(\sum_\sigma n^f_{\sigma}  -1\right)^2
\nonumber
\\
&=&
\sum_{\sigma} \left(\e_f  - \frac{\sigma}{2} H\right)
 f^\dagger_\sigma f_\sigma 
 + Un^f_{\uparrow}n^f_{\downarrow}
\nonumber \\
 \H_{U}&=&  \frac{U}{2}\left(\sum_\sigma n^f_{\sigma}  -1\right)^2
\end{eqnarray}
 consists of a single local state, which we will denote with $f$, with energy
$\e_f$ and Coulomb repulsion $U$,
coupled to a bath of conduction electrons with creation operators
$c^\dagger_{k\sigma}$ and energies $\e_{k\sigma}$. The local level is
subject to a Zeeman splitting in an external magnetic field $H$.
Note that the single-particle term of the impurity Hamiltonian
$\H_{imp}$ can be written in two different ways, i.e.~the last two
lines of Eq.~(\ref{eqn:25}) which allows for a conventional
interaction term -- last line of Eq.~(\ref{eqn:25}) -- or
non-interaction term containing the Hartree contribution and a
particle-hole preserving interaction term
$\H_{U}$ \cite{KrishWilWilson80a,KrishWilWilson80b}.  To obtain a continuous
spectral function  from the set of discrete $\delta$-functions
occurring in $G_{A,B}(z)$, the occurring $\delta(\w-\w_n)$ functions are
replaced by a Gaussian broadening on a logarithmic mesh 
 \begin{equation}
   \delta(\w -\w_n) \to \frac{e^{-b^2/4}}{b\w_n\sqrt{\pi}} \exp\left\{-\left(\frac{\ln(\w/\w_n)}{b}\right)^2\right\}
 \label{eqn:broadening}
 \end{equation}
where $b$ ranges typically between $0.6\le b <
1.2$\cite{SakaiShimizuKasuya1989,PetersPruschkeAnders2006,BullaCostiPruschke2007}.

The Fourier transformation of the Green function $G^r_{A,B}(t,t')$
with respect to $t$ obeys the equation of motion
\begin{equation}
  z G^r_{A,B}(z,t') =  \Tr{ \hat \rho(t') [ A, B ]_{s} }  +
  G^r_{[H,A],B}(z,t') 
\label{eqn:32}
\end{equation}
for any time $t'$ and a time-independent Hamiltonian $\H^f$. (Note
that a time-dependent  $\H^f(t)$ yields the usual integral equation,
and Eq.~(\ref{eqn:32}) would not hold.)

By setting $A=f_\sigma$ and $B=f^\dagger_\sigma$, Bulla et al.~derived
a simple but exact relation between two Green functions and the
correlation self-energy 
\cite{BullaHewsonPruschke98}
\begin{equation}
  \Sigma^U_\sigma(z,t') = U 
\frac{G^r_{f_\sigma n_{-\sigma},f^\dagger_\sigma }(z,t')}
{G^r_{f_\sigma ,f^\dagger_\sigma }(z,t')}
\label{eqn:self-energy-eom}
\end{equation}
which is used to express the retarded Green function as
\begin{eqnarray}
\label{equ:gf-eom}
  G^r_{f_\sigma ,f^\dagger_\sigma }(z,t') &=& 
\left[ z - \e_f
     - \frac{\sigma}{2} H - \Delta_\sigma(z)
     - \Sigma^U_\sigma(z,t')\right]^{-1} 
\komma
\nonumber\\
\Delta_\sigma(z) &=& \frac{1}{N}\sum_{k} \frac{V^2}{z-\e_{k\sigma}}
\punkt
\end{eqnarray}

We have calculated the Green functions $G^{r(NRG)}_{f_\sigma
  n_{-\sigma},f^\dagger_\sigma }(z,t')$ and $G^{r(NRG)}_{f_\sigma
  ,f^\dagger_\sigma }(z,t')$ in the steady-state limit $t'\to \infty$
and have obtained the 
physical Green function via the equation of motion (\ref{equ:gf-eom})
and (\ref{eqn:self-energy-eom}). 

As long as not otherwise stated, all
energies are measured in units of $\Gamma = \pi V^2 \rho(0)$, a
constant band width\cite{Wilson75} of $\rho(\w)=1/(2D)\Theta(D-|\omega|)$ is used with
$D/\Gamma=20$. The number of kept states after each NRG iteration
was $N_s=2000$.  The check the accuracy,  we calculated the sum-rule of the raw
NRG spectral function by integrating  the $\delta$-peaks analytically
and confirmed that for arbitrary parameters and number of states the
sum-rule for the steady-state spectral function is fulfilled within
machine precision of $10^{-15}$. The algorithm itself combines the
time-dependent 
NRG\cite{AndersSchiller2005,AndersSchiller2006} implementation with
the calculation of the sum-rule conserving spectral functions as
discussed elaborately in Ref.~\cite{PetersPruschkeAnders2006}.

\subsection{Particle-hole symmetry}

\subsubsection{External magnetic field $H=0$.}

\begin{figure}[tb]
  \centering


  \includegraphics[width=85mm,clip]{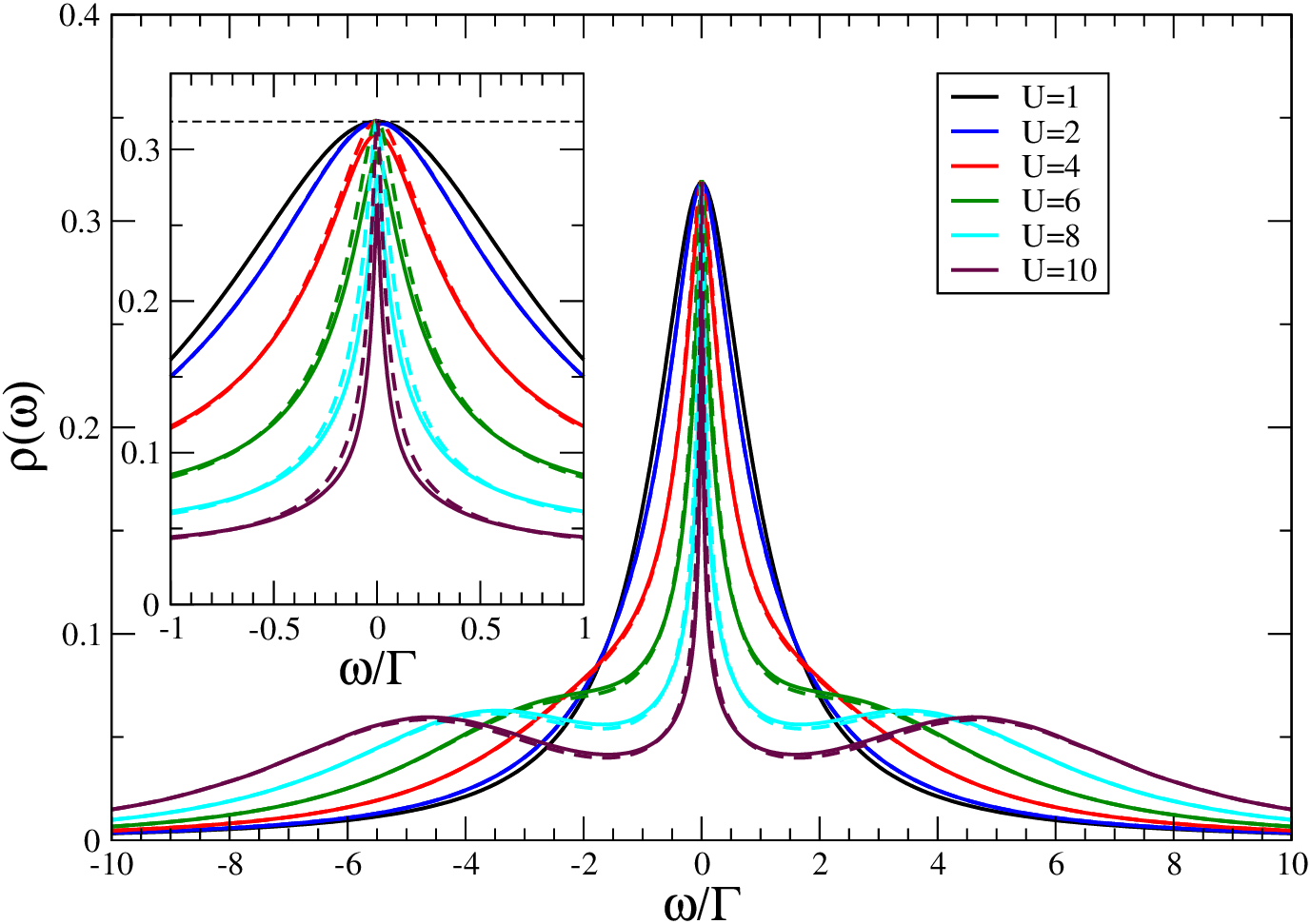}
  \caption{(color online) Comparison of the spectral function for the
    six different values of $U$ for the symmetric case
    $\e_f=-U/2$. The steady-state spectral function, obtain from
    switching $\H_U=0$ to a finite value is plotted as straight line,
    while the direct equilibrium
    calculation\cite{PetersPruschkeAnders2006} is given by a  dashed
    lines of same color for the same parameters. The inset shows the
    resonance in the 
    vicinity of the chemical potential. The dashed line in the inset
    indicated the unitary limit of $1/(\pi \Gamma)$.
    NRG parameters:  $\Gamma/D =\pi V^2\rho_0/D=0.05, 
    \Lambda=2, N_{S}=2000, b=0.6, T \to 0$.}
  \label{fig:2}
\end{figure}

In Fig.~\ref{fig:2}, the steady-state spectral functions for a
particle-hole symmetric regime are compared with the equilibrium
solution obtained directly from the standard NRG
procedure\cite{PetersPruschkeAnders2006}. In these calculations, the
Hartree term $U/2$ has been absorbed into $\H^i$. At time $t'=0$,
the Coulomb interaction $\H_U$ is switched on. An excellent agreement between the 
equilibrium NRG result (dashed lines) and the long-time limit of the
time-evolved spectral functions (solid lines) is found. The
non-interacting resonant-level spectral function centered around $\w=0$
evolves continuously into the Green function for a SIAM with finite
$U$. The inset in Fig.~\ref{fig:2} shows small 
deviations between the reference equilibrium spectra for $\H=\H^f$ and
the steady-state spectra obtained from the Fourier-transform of
Eq.~(\ref{eqn:gf-st}) in the Kondo regime. 
Note that 
the exponentially small Kondo scale not accessible to perturbation
theories in $U$ is always accounted for correctly within the NRG and,
therefore, in our algorithm by the  crossover to the
fixed-point spectrum  of $\H^f$\cite{KrishWilWilson80a,KrishWilWilson80b}.
With increasing values of
$U$ and fixed $\Lambda$, the peak height decreases from its
theoretical unitary limit of $1/(\pi \Gamma)$. The deviations are less
that 1\% for $U=2$ and increase to approximately 11\% for $U=10$. 
The correct low-energy scale\cite{KrishWilWilson80a,BullaCostiPruschke2007}
$T_K$ proportional to the width of the resonance  
at $\w=0$ emerges as well in the steady-state spectral functions.

\begin{figure}[tb]
  \centering

  \includegraphics[width=85mm,clip]{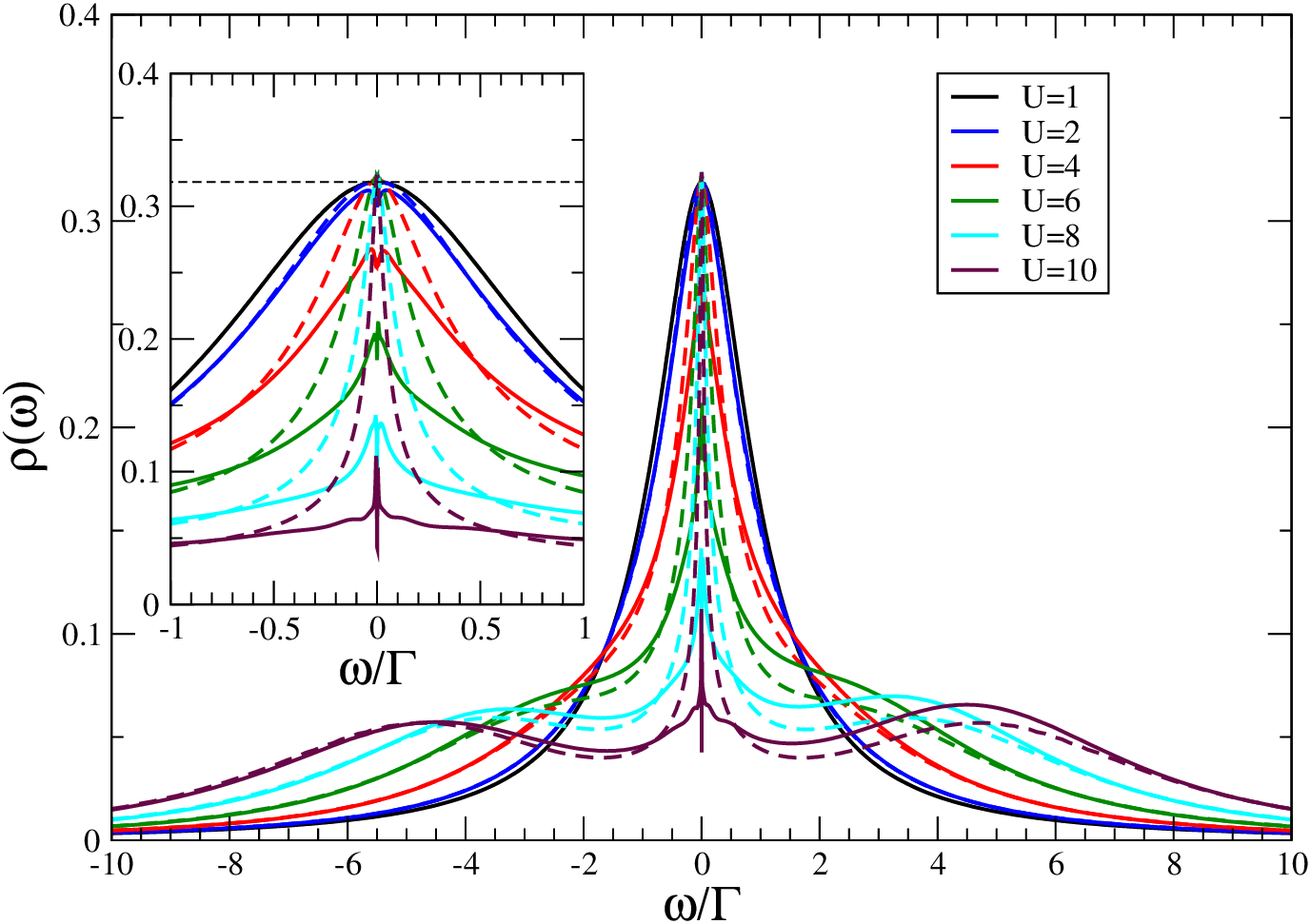}
  \caption{(color online) Comparison of the spectral function for the
    six different values of $U$ for the symmetric case
    $\e_f=-U/2$. The Hartree term $U/2$ is absent in Hamiltonian of
    $\H^i$, and the Coulomb interaction $H_U=  U n^f_{\uparrow}
    n^f_{\downarrow} = \H^f -\H^i$ is switched on at $t'=0$.  The
    steady-state spectral function  are plotted as
    solid  lines, while the direct equilibrium
    calculation\cite{PetersPruschkeAnders2006} yields the dashed lines
    for the same parameters. The colors (color online) are identical
    for the same values of $U$. The inset shows the resonance in the
    vicinity of the chemical potential. NRG parameters:  as in
    Fig.~\ref{fig:2}. 
    }
  \label{fig:3}
\end{figure}

\begin{figure}[tb]
  \centering
   \includegraphics[width=85mm,clip]{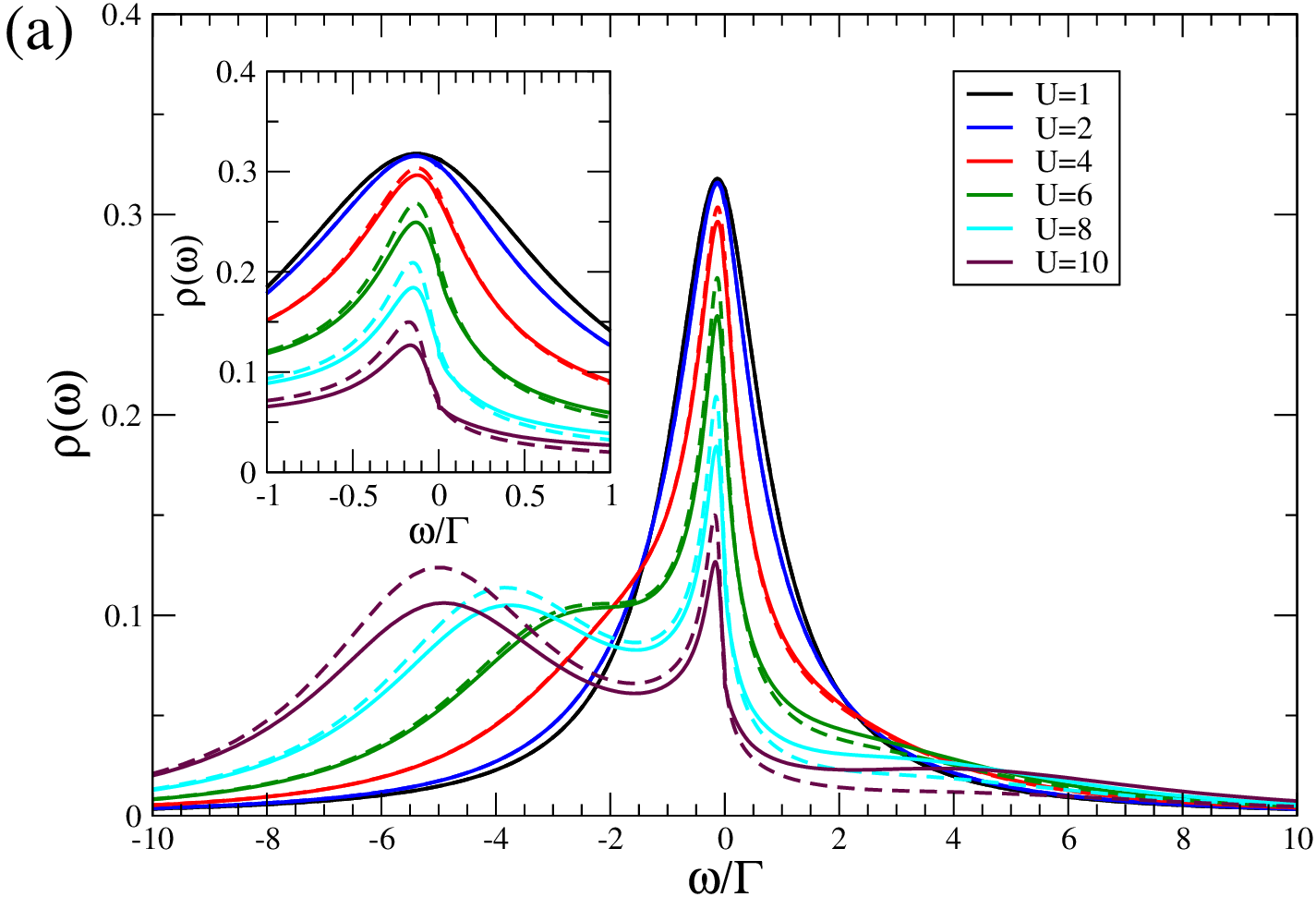}

   \includegraphics[width=85mm,clip]{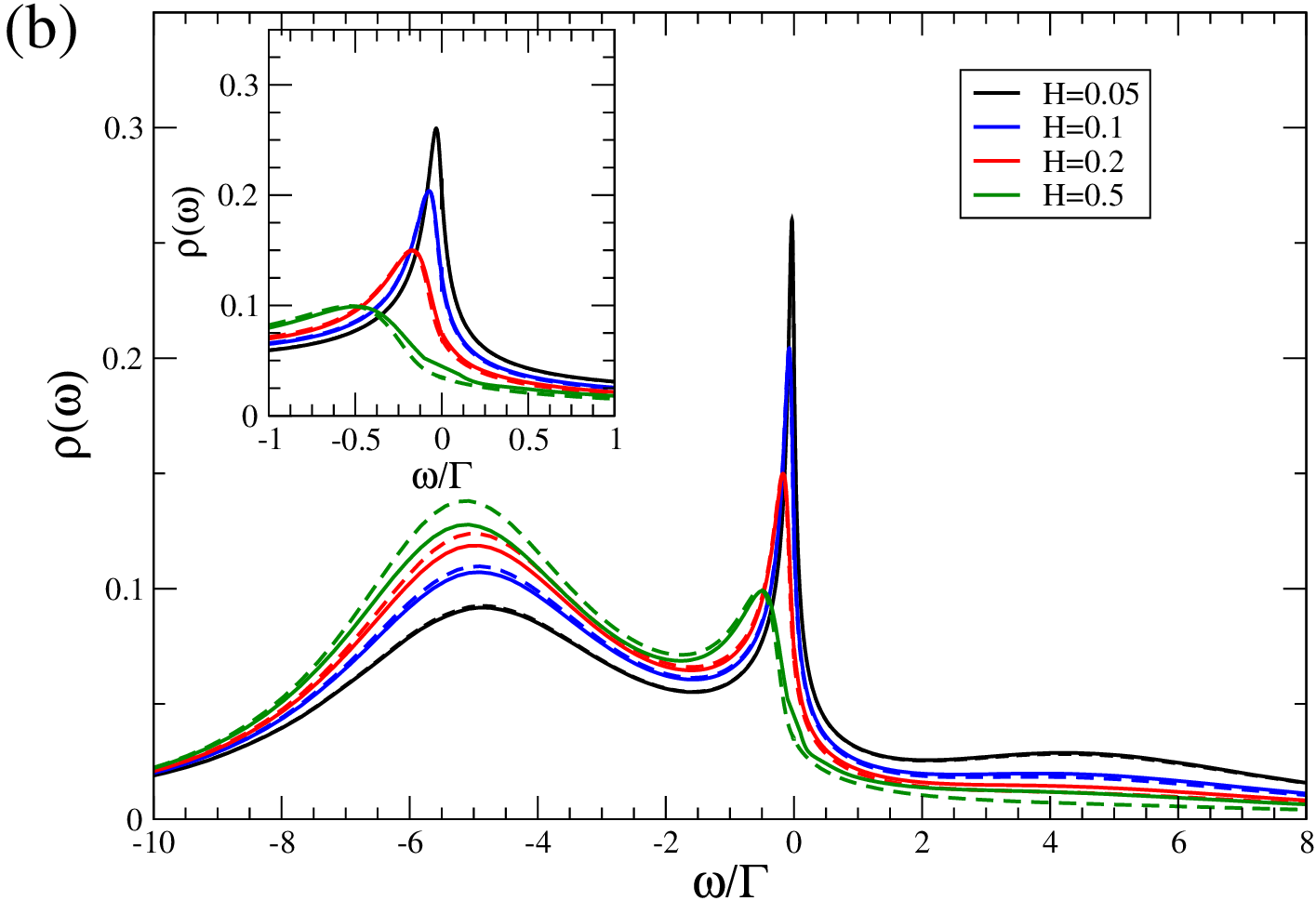}

   \caption{(color online) Comparison (a) of the majority spin spectral
    function for the     six different values of $U$ for the symmetric
    case      $\e_f=-U/2$ at a fixed finite magnetic field
    $H=0.2$. The color coding and NRG parameters are identical to 
    Fig.~\ref{fig:2}.    The steady-state spectral functions, obtain
    by switching $\H_U=0$ to a finite value are plotted as
    straight lines,   while the direct equilibrium
    calculation\cite{PetersPruschkeAnders2006} is given by the dashed
    lines with the same color for the same $\H^f$.
    In (b) $U/\Gamma=8$ and $\e_f/\Gamma=-4$ has been kept constant
    while the external magnetic field is switch on.
    The inset shows the resonance in the     vicinity of the chemical
    potential.
    NRG parameters:  as in Fig.~\ref{fig:2}.
    }
  \label{fig:finite-H}
\end{figure}

We investigated also the impact of the initial level position $\e_f^i$
onto the steady-state spectra. A different starting point for $U=0$
could be the traditional way of writing of the impurity
Hamiltonian  $\H_{imp} = \sum_{\sigma} \left(\e_f  - \frac{\sigma}{2}
  H\right) f^\dagger_\sigma f_\sigma   + U n^f_{\uparrow}
n^f_{\downarrow}$ which is identical to (\ref{eqn:h-imp}). Here, the
Hartree term $U/2$ is not absorbed into  the single-particle energy
and  the   Coulomb repulsion term $\H_U = U n^f_{\uparrow}
n^f_{\downarrow}$ is switched on at $t'=0$.

The results for this starting point are presented in Fig.~\ref{fig:3}.
The  steady-state spectra  show an  increasing deviation from the correct
thermodynamic equilibrium spectrum which remains pinned at $1/\pi$ for
all values of $U$ in accordance with the density of state sum
rule\cite{Langreth1966,AndersGreweLorek91}. All steady-state spectra
remain  particle-hole symmetric, guarantied by $\H^f$, and the high
energy feature are well reproduced. However, we observe deviations
from the correct Abrikosov-Suhl resonance (ASR) already for moderate
values of $U>2\Gamma$. For large values of $U$, the ASR is almost
absent in the  steady-state spectra.  

The difference can be understood in the following way. By absorbing
the Hartree term into the initial Hamiltonian $\H^i$, the average
impurity occupation $\expect{n_f}$ does not change with time. $\H^i$
and $\H^f$  will flow to the same strong-coupling fixed point for
$T\to 0$.  The excellent agreement between the equilibrium
reference spectrum and the steady-state spectrum  can be seen in
Fig.~\ref{fig:2}.   

In Fig.~\ref{fig:3}, however, we have started with a non-interacting
Hamiltonian which 
breaks particle-hole symmetry: the level position is located at
$\e_f = -U/2$. For increasing values of $U/\Gamma>1$, it corresponds
to a doubly occupied level as the starting configuration while the final
spectra must be particle-hole symmetric for $\e_f = -U/2$.  The
strong-coupling fixed point of $\H^i$ is characterized by an
additional marginal operator which is proportional to the strength of the
particle-hole symmetry breaking\cite{KrishWilWilson80b}. For energies 
larger than the characteristic energy scale $T_K$, a good agreement is
found for the high energy parts of the spectrum which is determined
mainly by the mean occupation. However, the low energy spectrum, which
contains the information on the many-body resonance, deviates
increasingly with increasing values of $U$ from the reference curve.

\subsubsection{Finite external magnetic field }

The particle-hole symmetry, present at $H=0$ is broken at a finite
magnetic field.
In Fig.~\ref{fig:finite-H}(a), a comparison is shown  between the
equilibrium  spectral functions (dashed lines) and $\rho(\w,t'\to
\infty)$ obtained after switching on a finite value of $U$ in a fixed
and finite magnetic field of $H=0.2$. The position and height of the many-body
resonance is  well reproduced. The small deviations for the equilibrium
values increase with increasing value of $U$. A shift in spectral
weight from negative to positive frequencies of the majority spectrum
at large values of $U$ indicates a  slight underestimation of the
spin-polarization for values of $U \ge 8$. Due to the total spin
conservation of the Hamiltonian, a relaxation of the total
magnetization is prohibited. This is the source of additional small
deviations\cite{AndersSchiller2005,AndersSchiller2006}  besides
discretization errors in the finite-size representation of the
infinitely large system.

Alternatively, we have  kept  $U$ fixed and switched on a
finite magnetic field $H$ at $t'=0$  as depicted in
Fig.~\ref{fig:finite-H}(b). Again, the equilibrium spectra is well
reproduced by $\rho(\w,t'\to \infty)$.

\begin{figure}[tb]
  \centering
  \includegraphics[width=85mm,clip]{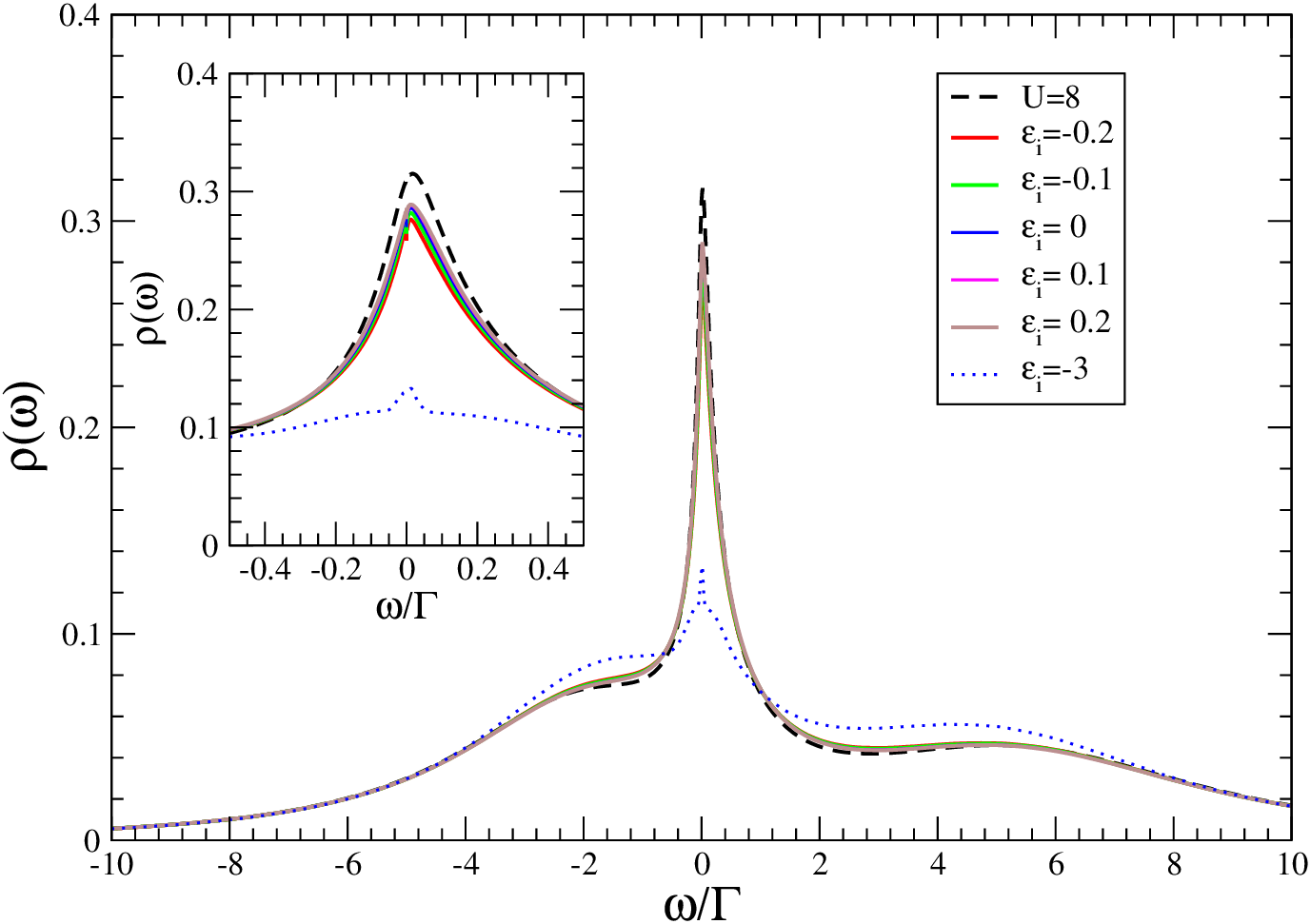}

  \caption{(color online) Influence of the initial value of the level
    position in $\H^i$ on the steady-state spectrum for a fixed value
    of $U=8$. The initial level position $\e_f$ has been set to 
    $\e^{i}_f/\Gamma= -3,-0.2,-0.1,0,0.1,0.2$. The black dashed line
    shows the equilibrium NRG spectra for the small parameters as $\H^f$.
    The inset shows the resonance in the vicinity of the chemical potential. 
    NRG parameters:  as in Fig.~\ref{fig:2}.
    }
  \label{fig:ph-asymmetic-ei-scan}
\end{figure}

\begin{figure}[tb]
  \centering

  \includegraphics[width=85mm,clip]{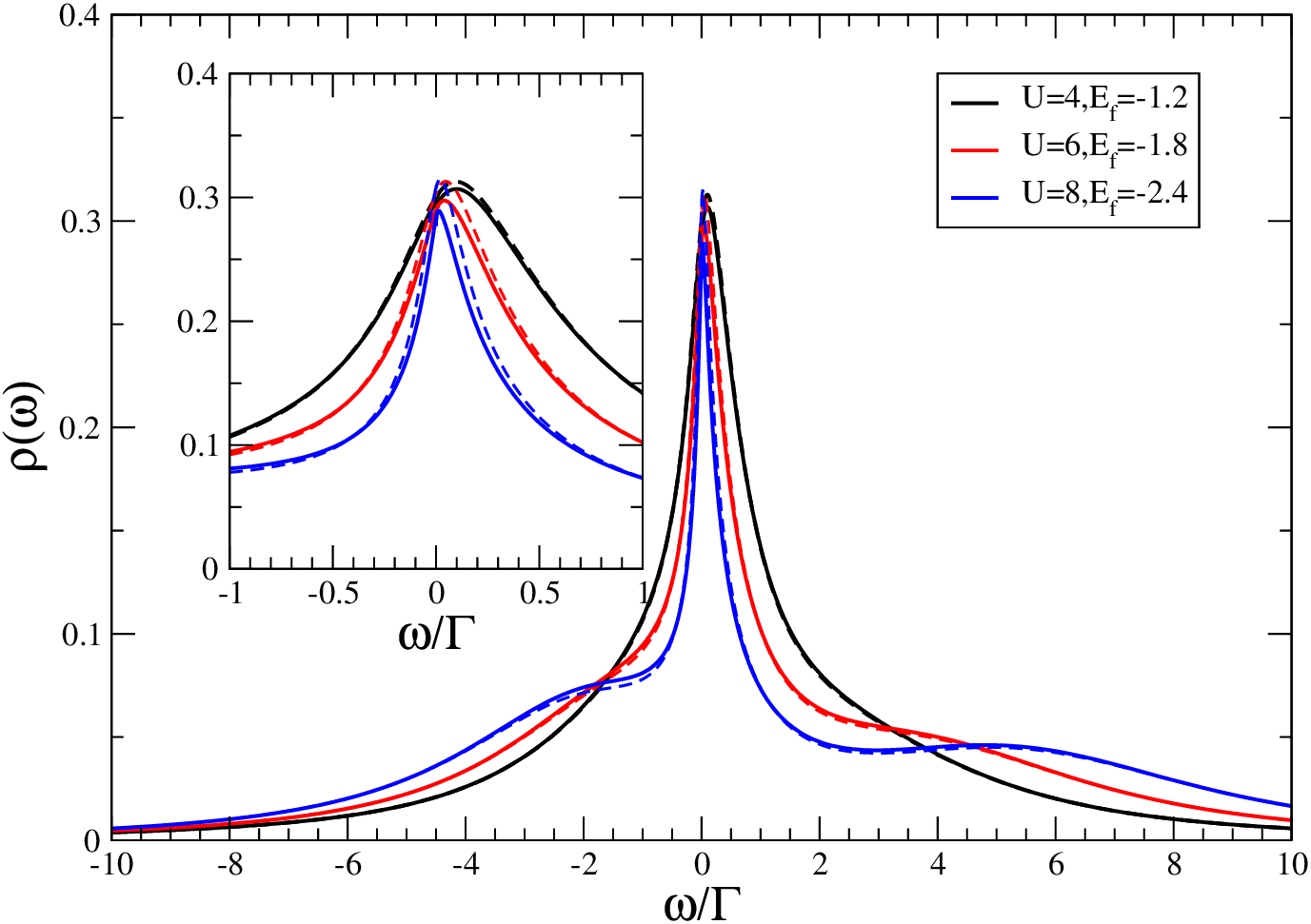}

  \caption{(color online) Comparison of the spectral function for the
    three different values of $U$ for the asymmetric case. 
    The initial level position $\e_f$ has been set to 
    $\e^{i}_f=0.235,0.21,0.175$ and $\e_f^f=-2.4$
    The inset shows the resonance in the 
    vicinity of the chemical potential. 
    NRG parameters:  as in Fig.~\ref{fig:2}.
    }
  \label{fig:ph-asymmetic}
\end{figure}

\begin{figure}[tb]
  \centering
  \includegraphics[width=85mm,clip]{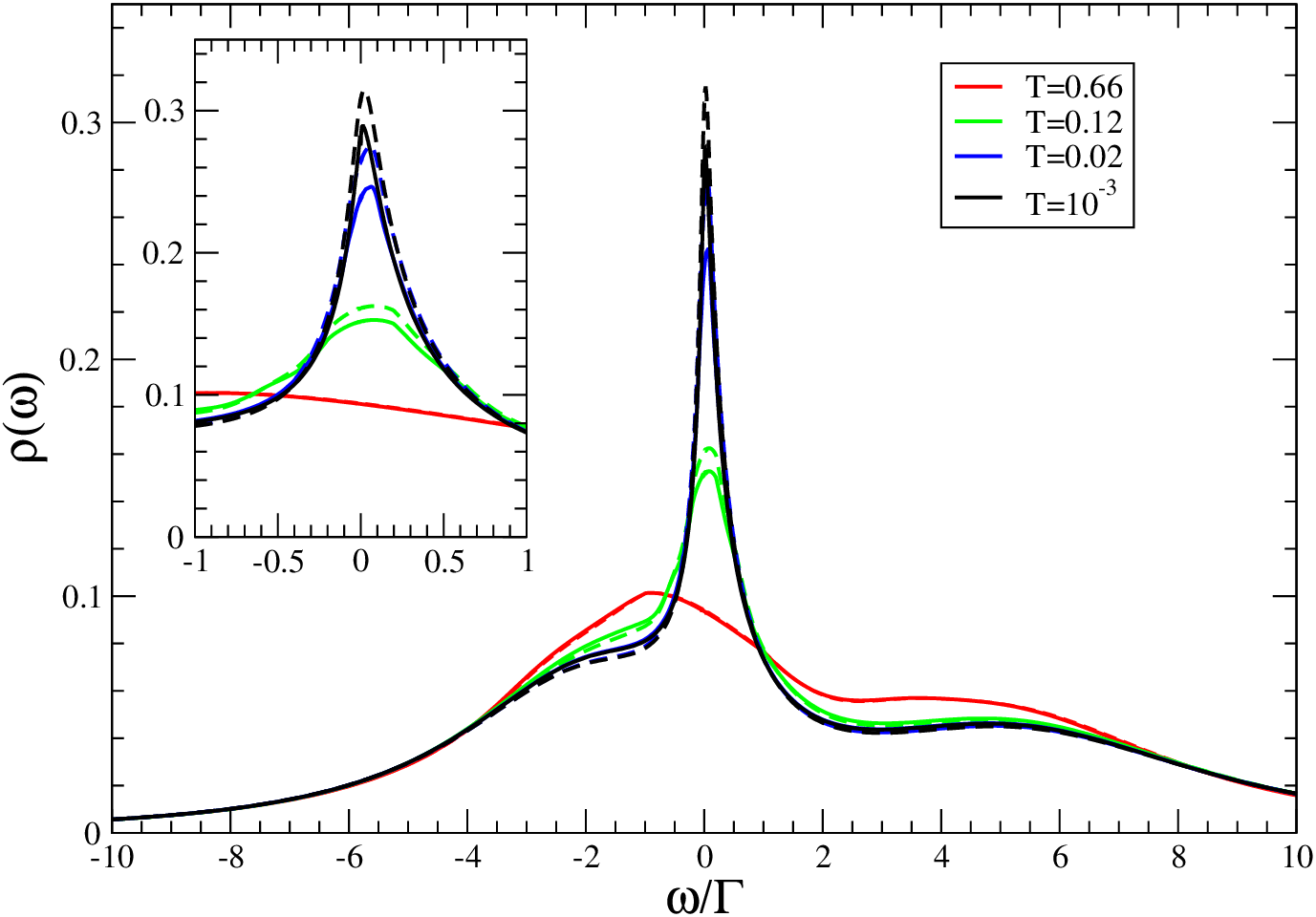}

  \caption{(color online) Comparison of the steady-state spectra
    (solid line) for   a fixed value of $U=8$ and $\e_f=-2.4$ evolved
    from $U=0$ and the thermodynamic equilibrium spectra (dashed line) 
    for different values of the temperature
    $T/\Gamma=0.66,0.12,0.02,10^{-3}$. The initial level position
    $\e_f$ has been set to $\e^{i}_f/\Gamma=0.175$. The black dashed line
    shows the equilibrium NRG spectra for the small parameters as $\H^f$.
    The inset shows the resonance in the vicinity of the chemical potential. 
    NRG parameters:  as in Fig.~\ref{fig:2}.
    }
  \label{fig:finite-T}
\end{figure}

\subsection{Particle-hole asymmetric regime}

The influence of the initial level position $\e_f^i$ on the
steady-state spectra is depicted in
Fig.~\ref{fig:ph-asymmetic-ei-scan} for  local particle-hole
asymmetric parameters $\e_f^f=-2.4$ and $U=8$. Again, we start 
initially with $U=0$. For variation of $\e_f^i$ 
which changes the level occupancy $n_f$  very moderately, the
steady-state spectral function shows only marginal changes. 
We observe a significant deviation from the equilibrium NRG spectral
function only for a large negative initial value of
$\e_f^i/\Delta=-3$, for which  the impurity is essentially doubly
occupied. Although the shape and position of the high-energy excitation
maxima are well reconstructed in this case, the strongly reduced
spectral weight of the low frequency resonance close to the chemical
potential requires additional spectral weight at high energies,
 a consequence of the sum-rule conserving algorithm. 

Particle-hole asymmetric spectral functions are displayed in
Fig.~\ref{fig:ph-asymmetic} for three different values of $U$. 
Here, we have chosen the non-interaction resonant level model $\H^i$
such that the low-temperature fixed point spectra is identical to the
one of $\H^f$.

Since the algorithm always evaluates the spectral function at a finite
temperature defined by $T_N \propto \Lambda^{-N/2}$ of the last NRG
iteration\cite{Wilson75,KrishWilWilson80a,KrishWilWilson80b,BullaCostiPruschke2007}
we can also track the temperature evolution of the spectra. For one set of
parameters used in Fig.~\ref{fig:ph-asymmetic}, such a temperature
evolution of the steady-state spectra is shown in
Fig.~\ref{fig:finite-T}. Dashed and solid lines of equal color
(color-online) correspond to the same
temperature. Fig.~\ref{fig:finite-T} clearly demonstrates that the
steady-state algorithm can be used for the temperature evolution of
spectral functions as well.

\section{Conclusion and Outlook}

We have presented a new algorithm to calculate non-equilibrium Green
functions $G(t,t')$ for quantum-impurity models.  It is derived using
the complete basis set for the Wilson NRG
chain\cite{AndersSchiller2005,AndersSchiller2006}. Therefore, the
spectral sum-rule is always fulfilled exactly, independent of the
number  $N_s$ of kept states after each NRG iteration.  We have shown
the algorithm for calculating equilibrium spectral
functions\cite{WeichselbaumDelft2007,PetersPruschkeAnders2006} is
included in our approach for the case of an unaltered Hamiltonian $\H^i=\H^f$.

We believe, that this algorithm will open new doors for theoretical
calculations of non-equilibrium quantum systems. In another 
publication\cite{AndersSSnrg2008}, we have applied our method to a
non-equilibrium problem for which the answer is not known a priori: an
open quantum system comprising of a quantum dot coupled to two leads
whose chemical potential difference   drives a current through this
interacting junction. Only for the non-interacting problem ($U=0$),
the exact solution is known\cite{Hershfield1993}. However, by switching on the
full Coulomb repulsion $\H_U$ at finite bias, the
steady-state non-equilibrium spectral function evolves from this
initially known solution. The steady-state currents through  an
interacting nano-device  is  accessible to the numerical
renormalization group method in the strong-coupling regime at finite
bias. This method has  the advantage that it is applicable to
any arbitrary coupling strength, magnetic field and temperature. In
contrast to perturbative approaches it allows the study of the
crossover from the weak-coupling regime at high temperatures to the
strong-coupling regime at low temperatures and finite bias.

In this paper, we have restricted ourselves to the relevant case of
switching on a finite Coulomb repulsion $U$ at $t'=0$. Focusing on the
steady-state limit $t'\to \infty$, we used the well 
studied equilibrium spectral functions of the SIAM as benchmark for
the steady-stated spectra obtained with our method. Since a closed
quantum impurity system will evolve into its thermodynamic
equilibrium\cite{DoyonAndrei2005}, if only $\H_{imp}+\H_I$ is changed,
the deviation between the steady-state and the equilibrium spectra serves
as a measure for the quality of the algorithm.

We have shown that the steady-state spectral functions agree
excellently  with the corresponding equilibrium spectra even at finite
magnetic field.  The absorbing of the Hartree term into the
non-interacting part of the Hamiltonian yields the best agreement
between the steady-state spectra and the equilibrium NRG spectra
directly obtained from $\H^f$.  The singly peaked spectrum of the
resonant level model evolves into the typical three peak structure
of the SIAM in the Kondo regime, with the lower and high frequency
peaks resulting from charge fluctuations and a narrow many-body Kondo
resonance emerging close to the chemical potential whose width is
proportional to the correct low energy scale.


\ack
We acknowledge discussions with N.~Andrei, R,.~Bulla,
G.~Czycholl, M. Jarrell,  Th. Costi, N.~Grewe, H.~Monien, A.~Millis,
T.~Novotny, J.~Kroha, Th.~Pruschke, A.~Schiller, P.~Schmitteckert,
A. Weichselbaum, J.~von Delft and the KITP for its  
hospitality, at which some of the work has been carried out. 
This research was supported in parts by the DFG  projects
AN 275/5-1 and AN 275/6-1
and by the National Science Foundation under Grant No. PHY05-51164
(FBA). We acknowledge supercomputer support by the NIC, Forschungszentrum
J\"ulich under project no.\ HHB000 (FBA). 


\section*{References}


\providecommand{\newblock}{}

\end{document}